\documentclass[]{revtex4}

\usepackage{graphicx}
\usepackage{dcolumn}
\usepackage{bm}
\usepackage{hyperref}
\usepackage{epsfig}

\begin{document}



\title{Ion-neutral chemistry at ultralow energies: \\ Dynamics of reactive collisions between laser-cooled Ca$^+$ ions and Rb atoms in an ion-atom hybrid trap}
\author{Felix H.J. Hall}
\author{Pascal Eberle}
\author{Gregor Hegi}
\affiliation{Department of Chemistry, University of Basel, Klingelbergstrasse 80, 4056 Basel, Switzerland}
\author{Maurice Raoult}
\author{Mireille Aymar}
\author{Olivier Dulieu}
 \email{olivier.dulieu@u-psud.fr}
\affiliation{Laboratoire Aim\'{e} Cotton, CNRS/Univ. Paris-Sud/ENS Cachan, B\^{a}t. 505, Campus d'Orsay, 91405 Orsay Cedex, France}
\author{Stefan Willitsch}
 \email{stefan.willitsch@unibas.ch}
\affiliation{Department of Chemistry, University of Basel, Klingelbergstrasse 80, 4056 Basel, Switzerland}


\begin{abstract}
\bigskip
\noindent Cold chemical reactions between laser-cooled Ca$^+$ ions and Rb atoms were studied in an ion-atom hybrid trap. Reaction rate constants were determined in the range of collision energies $ \langle E_{\text{coll}} \rangle / k_{\text{B}}=20$~mK-20~K. The lowest energies were achieved in experiments using single localized Ca$^+$ ions. Product branching ratios were studied using resonant-excitation mass spectrometry. The dynamics of the reactive processes in this system (non-radiative and radiative charge transfer as well as radiative association leading to the formation of CaRb$^+$ molecular ions) have been analyzed using high-level quantum-chemical calculations of the potential energy curves of CaRb$^+$ and quantum-scattering calculations for the radiative channels. For the present low-energy scattering experiments, it is shown that the energy dependence of the reaction rate constants is governed by long-range interactions in line with the classical Langevin model, but their magnitude is determined by short-range non-adiabatic and radiative couplings which only weakly depend on the asymptotic energy. The quantum character of the collisions is predicted to manifest itself in the occurrence of narrow shape resonances at well-defined collision energies. The present results highlight both universal and system-specific phenomena in cold ion-neutral reactive collisions.

\end{abstract}

\maketitle

\section{Introduction}

The recent development of hybrid traps for the simultaneous cooling and trapping of ions and atoms has opened up perspectives for the study of ion-neutral interactions and chemical reactions at ultralow energies \cite{grier09a, zipkes10a, schmid10a, hall11a, rellergert11a, ravi12a, hall12a, willitsch12a}. Current experiments utilize laser cooling of atomic ions \cite{grier09a, zipkes10a, schmid10a, hall11a, rellergert11a, churchill11a, ravi12a, sivarajah12a, hall13b} or sympathetic cooling of molecular ions \cite{hall12a} in radio frequency (RF) traps with the simultaneous magnetic \cite{zipkes10a}, optical-dipole \cite{schmid10a}, or magneto-optical trapping and cooling \cite{grier09a, hall11a, rellergert11a, ravi12a, hall12a} of neutral atoms. These developments have enabled the study of ion-neutral collisional processes down to collision energies corresponding to a few mK. In this domain (often referred to as the "cold'' regime), only a few partial waves contribute to the collisions, and resonance as well as radiative effects can become important \cite{cote00a, bodo02a, idziaszek09a, weiner99a, belyaev12a}.

A wealth of collisional and chemical phenomena has recently been observed in these systems, ranging from fast near-resonant homonuclear charge exchange  \cite{grier09a} and the effect of the electronic \cite{hall11a} and hyperfine state  \cite{ratschbacher12a} on the reaction rate, to the formation of molecular ions by radiative association \cite{hall11a, hall13b} and the sympathetic cooling of ions by atoms \cite{ravi12a, sivarajah12a}. Beyond atomic systems, molecular ions have also recently been introduced into a hybrid trap enabling the study of distinctly molecular effects in cold reactive collisions \cite{hall12a}. 

Despite the breadth of both experimental and theoretical investigations on cold ion-neutral interactions performed thus far, important problems relating to the dynamics of these processes are still unresolved. At ultralow energies, only chemical reactions without an activation barrier can occur. One of the key questions in this context is whether such processes can be understood in terms of an ``universal'' behaviour, i.e., whether they are only governed by long-range intermolecular forces \cite{dashevskaya03a, nikitin05a, gao10a, gao10b, gao11a}, so that they can be described within the framework of capture models based on, e.g., the statistical adiabatic channel model (SACM) developed by Quack and Troe \cite{quack75a, troe87a, dashevskaya03a}. However, for cold ion-neutral reactions few attempts to directly compare the results of experiment and theory have been made thus far \cite{rellergert11a, zhang09a, hall13b}. Such comparisons are necessary to validate reactive-scattering models and gain an in-depth understanding of the cold-collision dynamics observed in ion-atom systems.

In a previous letter \cite{hall11a} (in the following referred to as paper I), we have reported the first results of a combined experimental and theoretical investigation of cold reactive collisions between Ca$^+$ ions and Rb atoms in a hybrid trap. Three reactive processes were found to be in competition with each other: non-radiative charge transfer (NRCT) mediated by non-adiabatic couplings between the molecular states, radiative charge transfer (RCT) induced by emission of a photon to continuum states associated with lower lying potential energy curves, and radiative association (RA) of CaRb$^+$ molecular ions by emission of a photon to lower-lying bound vibrational levels. A $>$100 times faster reaction rate was observed in the excited Ca$^+[(4p)~^2P_{1/2}]$+Rb$[(5s)~^2S_{1/2}]$ reaction channel in comparison to the lowest channel Ca$^+[(4s)~^2S_{1/2}]$+Rb$[(5s)~^2S_{1/2}]$. This observation was rationalized in terms of the dynamics occurring in a network of avoided crossings between potential energy curves (PECs) associated with the excited entrance channel and charge-transfer states which provide multiple pathways for NRCT, RCT and RA.

In the present paper, we present a detailed characterization of the dynamics of cold reactive collisions in the Ca$^+$+Rb system via a  comparison of the results of quantum-chemical and quantum-scattering calculations with the experimental data. The current work focuses on the theoretical modelling of the radiative processes, the study of the dependence of the rate constant on the collision energy and the rationalization of the observed CaRb$^+$:Rb$^+$ product branching ratios. To address these questions, the experiments have been extended to average collision energies $\langle E_{\text{coll}}\rangle /k_{\text{B}}\geq$20~mK by studying reactions of single Ca$^+$ ions. Calculations of radiative cross sections for ground state collisions have been performed at collision energies from $<$1~mK up to 70~K, enclosing the energy range accessed in the experiments. Theoretical predictions for the rate constants were obtained from the cross sections and realistic velocity distributions of the ions were extracted from molecular dynamics (MD) simulations of ion Coulomb crystals. The product branching ratios were estimated from a comparison of experimental mass spectra of the ions with MD simulations. By comparing theory and experiment, we show that the rates of reactive collisions in the present system can be understood in terms of a classical capture of the reaction partners induced by long-range interactions followed by reactive dynamics at short range. Whereas the former governs the energy dependence of the reaction rates, the latter is only weakly dependent on energy, but determines the absolute value of the rate constants. The present results highlight universal as well as system-specific aspects of cold ion-neutral collisions. 

The present article is organized as follows. In Sec. \ref{methods}, we give a detailed account of our experimental and theoretical procedures. Experimental results for the energy variation of the rate constants and product branching ratios are presented in Sec. \ref{results} and interpreted on the basis of computed potential energy curves and cross sections. The final discussion in Sec. \ref{conclusions} highlights the main characteristics of the ion-neutral reactive collisions in the cold regime.

\section{Methods}
\label{methods}

\subsection{The ion-atom hybrid trap}
\label{hybridtrap}

\begin{figure}[t]
\begin{center}
\epsfig{file=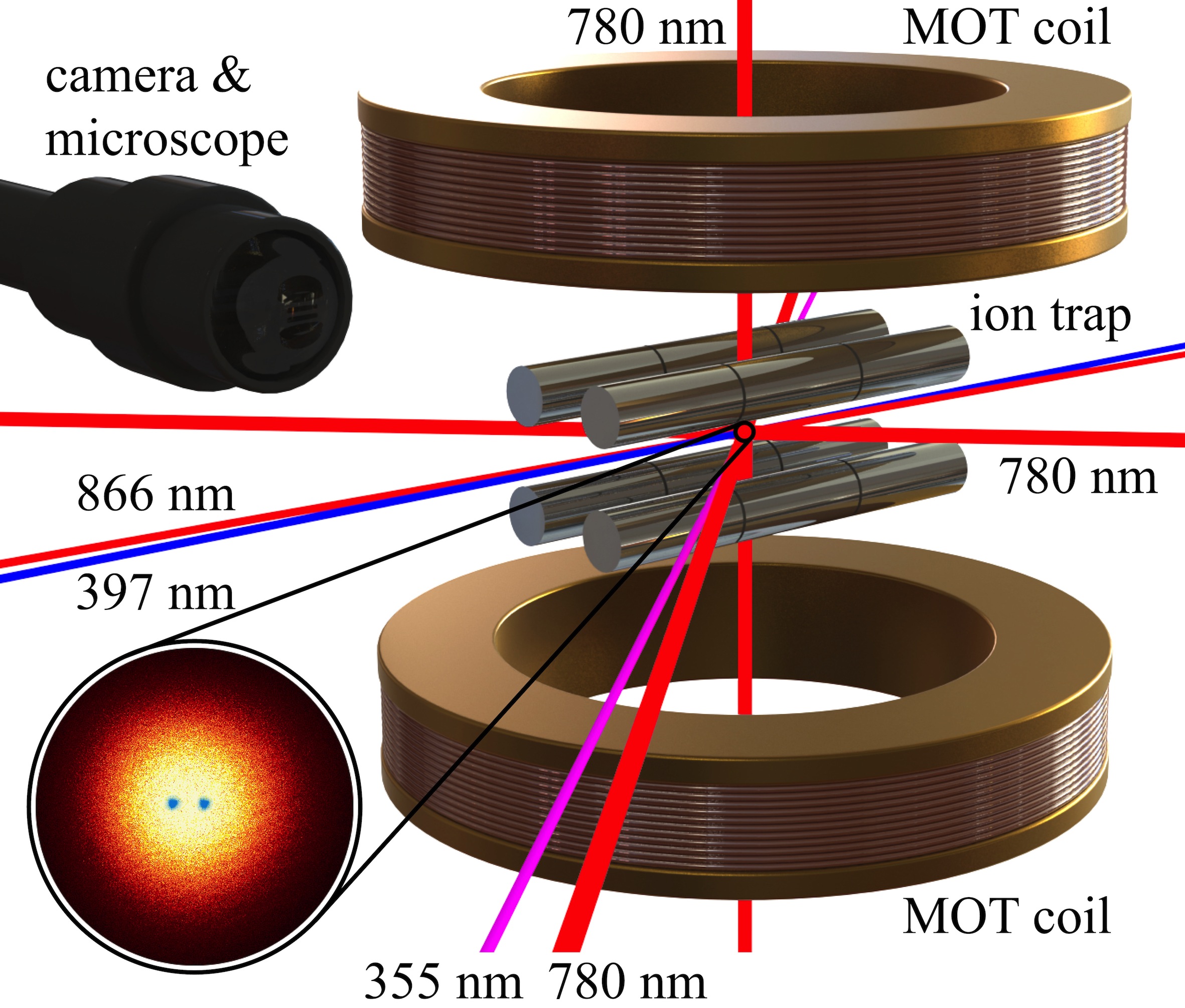,width=0.5\textwidth}
\end{center}
\caption{\label{setup} Schematic of the experimental setup. The inset shows a string of two laser-cooled Ca$^+$ ions (blue) immersed in a cloud of laser-cooled Rb atoms (yellow-red). See text for details.}
\end{figure}

The experimental apparatus for the simultaneous trapping of laser-cooled Ca$^+$ ions and ultracold Rb atoms has briefly been described in paper I \cite{hall11a} and a detailed description is given here. Ca$^+$ ions were produced by non-resonant photoionization from an atomic beam passing through the center of a linear radiofrequency (RF) ion trap \cite{willitsch12a}. Following the design of Ref. \cite{drewsen00a}, the linear RF trap consists of four segmented cylindrical electrodes with radius $r=4.0$~mm evenly distributed on a circle with an inscribed radius $r_0=7.3$~mm. The spacing between the electrodes was chosen to be wider than in an ideal linear quadrupole ($r_0\approx r$ \cite{gerlich92a}) to provide optical access for the Rb cooling and trapping laser beams (see below). Numerical simulations showed that the deviation from an ideal quadrupolar potential was only slight ($\approx 6\%$ within 5~mm of the central trap axis). Confinement of the ions in the plane perpendicular to the trap axis was achieved by applying RF voltages $V_{\text{RF}}(t)=V_{\text{0,RF}}\cos(\Omega_{\text{RF}}t)$ with opposite polarities across adjacent electrodes \cite{willitsch12a}. Typical RF amplitudes and frequencies used were $V_{\text{0,RF}}=280$~V and $\Omega_{\text{RF}}=2\pi\times3.1$~MHz, respectively. Trapping in the axial direction was achieved by applying static voltages $V_{\text{END}}=1-10$~V to the endcap segments of the electrodes \cite{willitsch12a}.

Laser cooling of the Ca$^+$ ions to secular temperatures $T_{\text{sec}}\leq 10$~mK was achieved using two laser beams at 397~nm and 866~nm pumping the $(4s)~^2S_{1/2}\rightarrow (4p)~^2P_{1/2}$ and $(3d)~^2D_{3/2}\rightarrow (4p)~^2P_{1/2}$ transitions, respectively. Under these conditions, the ions localize in the trap to form Coulomb crystals \cite{willitsch08a, willitsch12a} which were imaged by collecting the spatially-resolved laser-cooling fluorescence using a microscope (magnification $10\times$) coupled to a charge-coupled device (CCD) camera.

A magneto-optical trap (MOT) \cite{raab87a} for $^{87}$Rb was set up with two water-cooled solenoids enclosing the ion trap in vacuum (see Fig. \ref{setup}) to generate a quadrupolar magnetic field with a gradient of 20~G/cm. The coil assembly was mounted on a {\it xyz} precision translational stage enabling the precise mechanical alignment of the MOT with the ion trap without the need for magnetic compensation fields. Six laser beams at a wavelength of 780~nm intersected at the center of the quadrupolar magnetic field in an optical-molasses configuration for the cooling and magneto-optical trapping of $^{87}$Rb atoms on the $(5s)~^2S_{1/2}, F=2\rightarrow (5p)~^2P_{3/2},F=3$ transition. A fraction $\approx5\%$ of the laser beam intensity was frequency-shifted by 6.57~GHz using an electro-optical modulator (EOM) to enable repumping of the population from the $(5s)~^2S_{1/2}, F=1$ hyperfine level. The Rb atoms were captured from background vapour released by a getter source \cite{rapol01a}. The number densities (typically $n=1\times10^{9}$~cm$^{-3}$) and temperatures of the ultracold atoms in the MOT ($T=150-200~\mu$K) were determined using standard fluorescence-measurement and time-of-flight methods, respectively \cite{wieman95a, shah07a, pradhan08a}. In our experiments, the MOT- and ion-cooling laser beams were alternately blocked by a mechanical chopper operating at 1000~Hz to prevent photoionization of Rb in the $(4p)~^2P_{3/2}$ state by the 397~nm laser \cite{hall11a}.

In order to study the energy dependence of collisional processes in the hybrid trap, it was imperative to precisely determine the kinetic energies of the collision partners. In the present study, the collision energies were governed by the kinetic energies of the ions ($E_{\text{kin}}/k_{\text{B}}\geq10$~mK) which in turn were dominated by their fast oscillating motion driven by the RF fields (micromotion) \cite{gerlich92a, berkeland98a}. The energy stored in the micromotion scales quadratically with the distance of the ions from the central trap axis \cite{willitsch12a, berkeland98a} and only vanishes for ions precisely located on the axis. In the present study, Coulomb crystals containing up to hundreds of ions were used. In these large crystals, the ions are localized in a volume around the trap axis and exhibit a broad kinetic-energy distribution which was determined from molecular dynamics (MD) simulations (see Refs. \cite{bell09a, hall13b} and Sec. \ref{md} below). Experimentally, a precise centring of the Coulomb crystals on the trap axis (``axialization'') is mandatory to enable a reliable determination and control of the ion-kinetic-energy distributions. The methods employed to achieve the axialization of the Coulomb crystals and their maximum overlap with the cloud of ultracold atoms are detailed in Appendices \ref{ax} and \ref{apmot}, respectively.

Because both the ions and atoms were laser cooled, a substantial fraction of the collisions occurred with species in electronically excited states. Establishing the excited-state fractions of both species was necessary for the determination of state-dependent rate constants from the measurements \cite{hall11a}. For the laser-cooled Rb atoms, the population in the excited state of Rb was determined using standard techniques (see, e.g., Refs. \cite{wieman95a, rapol01a, shah07a}). The methods used for determining the Ca$^+$ populations are detailed in Appendix \ref{appop}.

\subsection{Mass spectrometry}
\label{massspec}

Because of the deep potential of the ion trap (the potential depth is 3-4 eV depending on the electrode potentials), the ionic reaction products remained trapped and were sympathetically cooled into the Coulomb crystal. The chemical identity of the product ions was established using resonant-excitation (RE) mass spectrometry \cite{drewsen04a, roth07a}. Motional excitation of the different ion species was achieved by the application of an additional RF field to the trap electrodes. The excitation couples to the Ca$^+$ ions via the Coulomb interaction resulting in a modulation of their laser-cooling fluorescence. The motional heating causes a dislocation of the Ca$^+$ crystal which was detected by monitoring the increase of ion fluorescence in a spatial region close to its equilibrium position. Resonant-excitation mass spectra were obtained by recording the fluorescence intensity in the detection region as a function of the excitation frequency. As discussed in Ref. \cite{roth07a}, a quantitative interpretation of the features in the RE mass spectra is difficult owing to their strong dependence on experimental parameters such as drive amplitude, scan direction, scan speed, and the crystal size and composition. The relative intensities of the peaks in the spectra can therefore only serve as a qualitative indication of the abundance of different ion species in the Coulomb crystal.

\subsection{Molecular-dynamics simulations}
\label{md}

MD methods were used to simulate the laser-cooling fluorescence images and RE mass spectra of the multi-component Coulomb crystals obtained in the present experiments. Simulated images were generated from ion trajectories obtained by solving the classical equations of motion of the ions in the trap under the action of laser cooling as explained in Refs. \cite{willitsch08b, bell09a}. A faithful simulation of a resonant-excitation experiment including the slow frequency sweep of the drive voltage over the experimental time scale (several minutes) was computationally not feasible. Instead, following Ref. \cite{roth07a} the mass spectra were simulated by Fourier-transforming the time-dependent ion kinetic energies after displacing the Coulomb crystals from the trap axis in the radial direction and relaxing their positions. This approach gives access to the same physical observable probed experimentally, i.e., the frequency spectrum of the ion crystal, and qualitatively reproduces the intensities of the peaks observed in the spectra \cite{roth07a}. To minimize computing time, these simulations were performed using the pseudopotential approximation for the trap \cite{willitsch12a} and an isotropic friction force for laser cooling. Moreover, an empirical background subtraction function of the form $I=a(1-\exp(-f/b))^c$ (where $I$ is the signal intensity, $f$ the motional frequency and $a$, $b$ and $c$ are adjustable parameters) was applied to remove spurious low-frequency components from the spectra resulting from chaotic ion motions during the relaxation of the Coulomb crystal in the simulation. 

\subsection{Scattering calculations}
\label{scattcalc} 

In the present theoretical treatment of radiative decay during collisions between ground-state Ca$^+$ and Rb species, the entrance channel is associated with a single adiabatic Born-Oppenheimer (ABO) state which is an excited electronic state of the CaRb$^+$ molecular ion. Spontaneous emission of a photon leads to the population of the ground electronic state of CaRb$^+$ whose ABO correlates at long distance with the lowest Ca+Rb$^+$ asymptote (see Fig. \ref{pec}). One can distinguish two processes according to the emitted photon energy and resulting products. Either the initial state decays to the dissociation continuum of the ground electronic state forming Ca and Rb$^+$ in their respective ground states (RCT), or a CaRb$^+$ molecular ion is formed by populating a bound vibrational level of the ground electronic state (RA). Hence, the spontaneous-emission cross-section can be formulated as the sum of two contributions \cite{zygelman88a, gianturco97a}:
\begin{eqnarray}
\sigma^{RCT}(\epsilon_{i})&=p&\frac{8\pi^2}{3c^3}\frac{\omega^3_{i,f}}{k_i^2} \sum_{J=0}^{\infty} \int_{0}^{\epsilon_{f}^{max}}\left( J | \langle J-1, \epsilon_{f}|D(R)|\epsilon_{i},J\rangle |^{2}\right. \nonumber \\
&&\left.+(J+1) |\langle J+1, \epsilon_{f}|D(R)|\epsilon_{i},J\rangle |^{2}\right) d\epsilon_{f}
\label{csrct}
\end{eqnarray}
and
\begin{eqnarray}
\sigma^{RA}(\epsilon_{i})&=p&\frac{8\pi^2}{3c^3}\frac{1}{k_i^2} \sum_{J=0}^{\infty} \sum_{v=0}^{v_{max}}\left( \omega^3_{iJ,v(J-1)} J | \langle J-1,v|D(R)|\epsilon_{i},J\rangle |^{2}\right. \nonumber \\
&&\left.+ \omega^3_{iJ,v(J+1)}(J+1)|  \langle J+1, v|D(R)|\epsilon_{i},J\rangle |^{2}\right).
\label{csra}
\end{eqnarray}
In Equations (\ref{csrct}) and (\ref{csra}), all quantities are expressed in atomic units. $p$ is the statistical weight of the entrance channel, $\omega$ is the angular frequency of the emitted photon and the ket $|\epsilon,J\rangle $ represents a partial wave component of an energy-normalized continuum wave function of the two nuclei corresponding to an energy $\epsilon$ at infinite internuclear separation. $\epsilon_{i}$ is the kinetic energy in the incoming channel $\epsilon_{i}=E-V_i(\infty)$, $k_{i}=\sqrt{2\mu \epsilon_{i}}$ is the associated wave number, $\mu$ is the reduced mass of the system and $\epsilon_{f}$ denotes the kinetic energy in the exit channel $\epsilon_{f}=E-V_f(\infty)-\hbar \omega_{i,f} $. If $J$ is the total angular momentum in the entrance channel, then $J'=J\pm1$ are the two possible allowed total angular momenta in the exit channel. In Eq. (\ref{csrct}), $\langle J', \epsilon_{f}|D(R)|\epsilon_{i},J\rangle$ is the transition dipole moment (TDM) between two energy-normalized continuum wave functions
\begin{equation}
\langle J', \epsilon_{f}|D(R)|\epsilon_{i},J\rangle=\int_{0}^{\infty} F_{J'}^{f}(\epsilon_{f},R) D(R) F_{J}^{i}(\epsilon_{i},R)~ dR
\label{dipcc}
\end{equation}
where $D(R)$ is the electronic dipole moment function. The continuum wave functions behave like
\begin{equation}
F_{\ell}(\epsilon,R)\sim\sqrt{\frac{2\mu}{\pi k}} \sin(kR-\ell \frac{\pi}{2}+\delta_{\ell}) 
\end{equation}
at large distance. In Eq. (\ref{csra}), the final state $|v,J\rangle $ is a discrete, bound rovibrational level of the ground electronic state of the CaRb$^+$ molecular ion. Accordingly, the transition dipole moment becomes
\begin{equation}
\langle J', v|D(R)|\epsilon_{i},J\rangle=\int_{0}^{\infty} \chi_{vJ'}^{f}(R) D(R) F_{J}^{i}(\epsilon_{i},R)~ dR.
\label{dipcb}
\end{equation}
Here, $\chi_{v,J}$ is a rovibrational wave function of the molecule normalized to unity. 

Our calculations cover an entrance-channel energy range from 10$^{-5}$~cm$^{-1}$ to 50~cm$^{-1}$. As a consequence,  up to 200 partial waves have to be taken into account in the calculations to obtain converged cross sections. The high $J$ values and low energies render it necessary to propagate the continuum wave function in the entrance channel to very large values of the internuclear distance $R\approx20000$~a.u. to ensure that the classically-allowed region situated at long range behind the centrifugal barrier is reached. Moreover, in the integral over the exit-channel energy appearing in Eq. (\ref{csrct}), an upper limit $\epsilon_{f}^{max} $=1000 cm$^{-1}$ and an integration step size $d\epsilon_{f}$= 1cm$^{-1}$ have been used to ensure convergence. In order to simplify the calculations and decrease the computation time, it has been assumed that the $J$ to $J\pm1$ TDMs are equal in both Eqs.(\ref{csrct}) and (\ref{csra}).

Previous theoretical studies of non-radiative charge transfer occurring in the Ca$^+(4s)$+Rb$(5s)$ channel \cite{belyaev12a} have predicted a large number of shape resonances in the reaction cross section within the energy range relevant for the present study. These resonances are caused by the trapping of the scattering wave function in a state bound behind the centrifugal barrier at well-defined collision energies. The resonances can exhibit widths as narrow as $10^{-4}$~cm$^{-1}$. Therefore, in order to reliably locate narrow resonances in the radiative cross sections a very small energy-step size has to be used in the calculations. Because of the broad energy range considered in the present work which spans more than six orders of magnitude, calculations with such a small step size would have required a prohibitive amount of computing time. Instead, the Milne phase-amplitude method was employed which yields a good estimation of resonance energies \cite{milne30a,korsch77a}.

\section{Results and discussion}
\label{results}

\subsection{Theoretical potential energy curves and cross sections}
\label{cs}

\begin{figure}[t]
\begin{center}
\epsfig{file=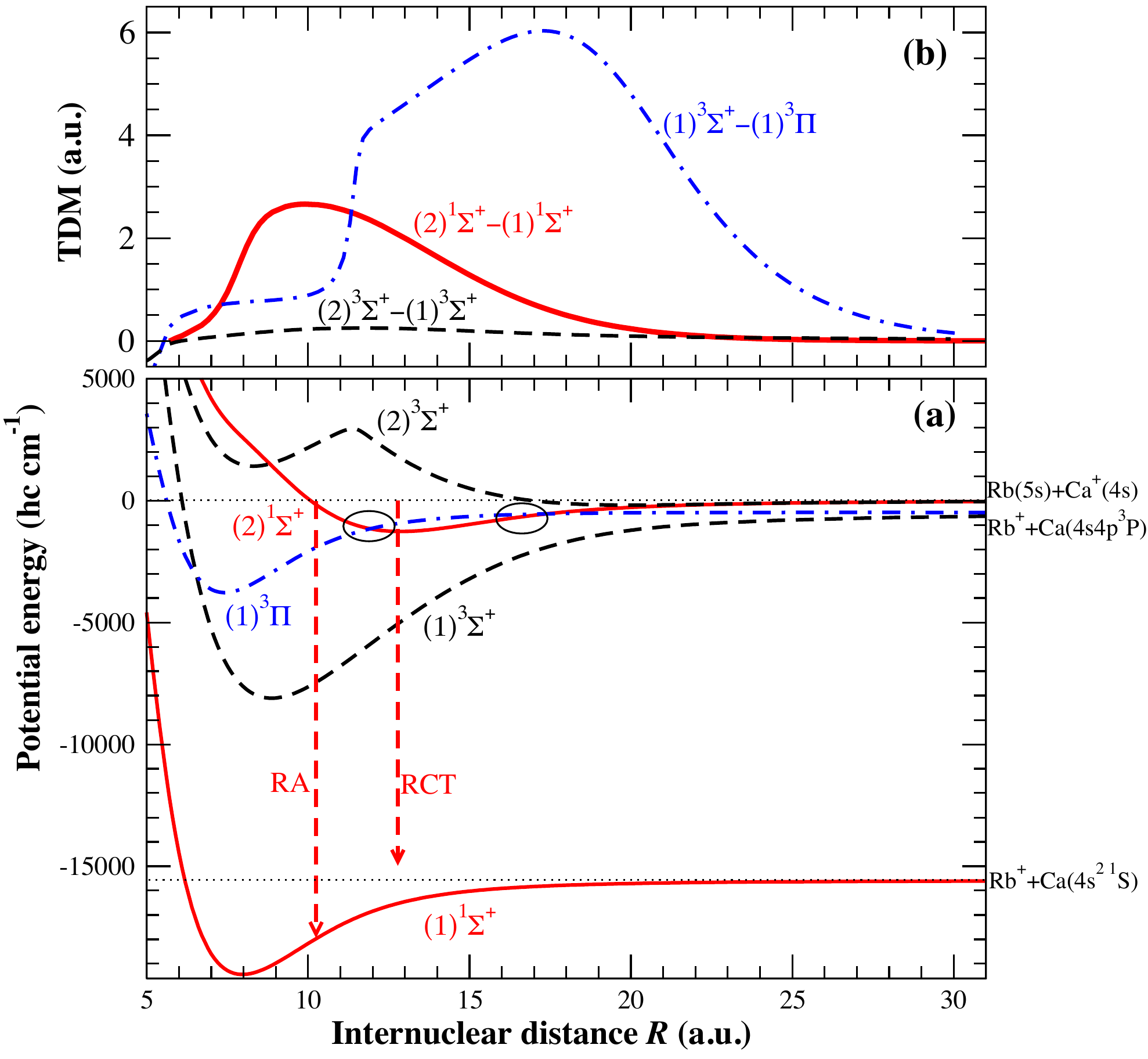,width=0.8\columnwidth}
\end{center}
\caption{\label{pec} (a) Computed non-relativistic potential energy curves (PECs) and (b) transition dipole moments (TDMs) relevant for the present study. Downwards arrows illustrate the radiative association (RA) and the radiative charge transfer (RCT) processes between the lowest singlet states. Circles indicate the crossing points between the $(2)^1\Sigma^+$ PEC associated with the lowest accessible collision channel Rb$(5s)$+Ca$^+(4s)$ and the $(1)^3\Pi$ PEC around which strong non-adiabatic interactions caused by spin-orbit coupling lead to non-radiative charge transfer (NRCT) \cite{tacconi11a}. }
\end{figure}

In Figure \ref{pec}, we display the potential energy curves correlated with the three lowest asymptotes of the (CaRb)$^+$ complex as well as the TDMs between the corresponding electronic states. The PECs have been calculated following the method of Ref. \cite{aymar05a, aymar11a} using effective core potentials (ECP) \cite{durand74a} representing the Rb$^+$ and the Ca$^{2+}$ ionic cores including core-polarization potentials (CPP) \cite{mueller84a} to simulate core-valence electronic correlation. Thus, only the two valence electrons are treated explicitly, allowing for a full configuration interaction (FCI) calculation performed in the configuration space generated by the large Gaussian basis sets reported in Refs. \cite{aymar11a,bouissou10a}. The fine structure of the calcium ion and of the rubidium atom are neglected here as justified below.

In the lowest collision channel accessed in the experiments (Ca$^+(4s)$+Rb$(5s)$), the collision between Ca$^+$ and Rb proceeds through two possible states, which were assumed to be statistically populated in the absence of polarization of the Rb atoms in the experiment:
\begin{itemize}
\item The almost purely repulsive $(2)^3\Sigma^+$ state (with a statistical weight $p=3/4$) in which the system is prevented from undergoing inelastic collisions. The magnitude of the TDMs (Fig. \ref{pec} (b)) and the small energy difference with the lower-lying triplet states suggest that the radiative emission from this state is negligible (recall that the RA and RCT cross sections scale with $\omega^3$, see Eqs. (\ref{csrct}) and (\ref{csra})).
\item The attractive $(2)^1\Sigma^+$ state (with a statistical weight $p=1/4$) which represents the main channel for inelastic collisions. As indicated in Fig. \ref{pec} (a), significant radiative emission is expected resulting in the formation of a molecular ion by RA or a Rb$^+$ ion and a Ca atom by RCT.
\end{itemize}

\begin{figure}[t]
\begin{center}
\epsfig{file=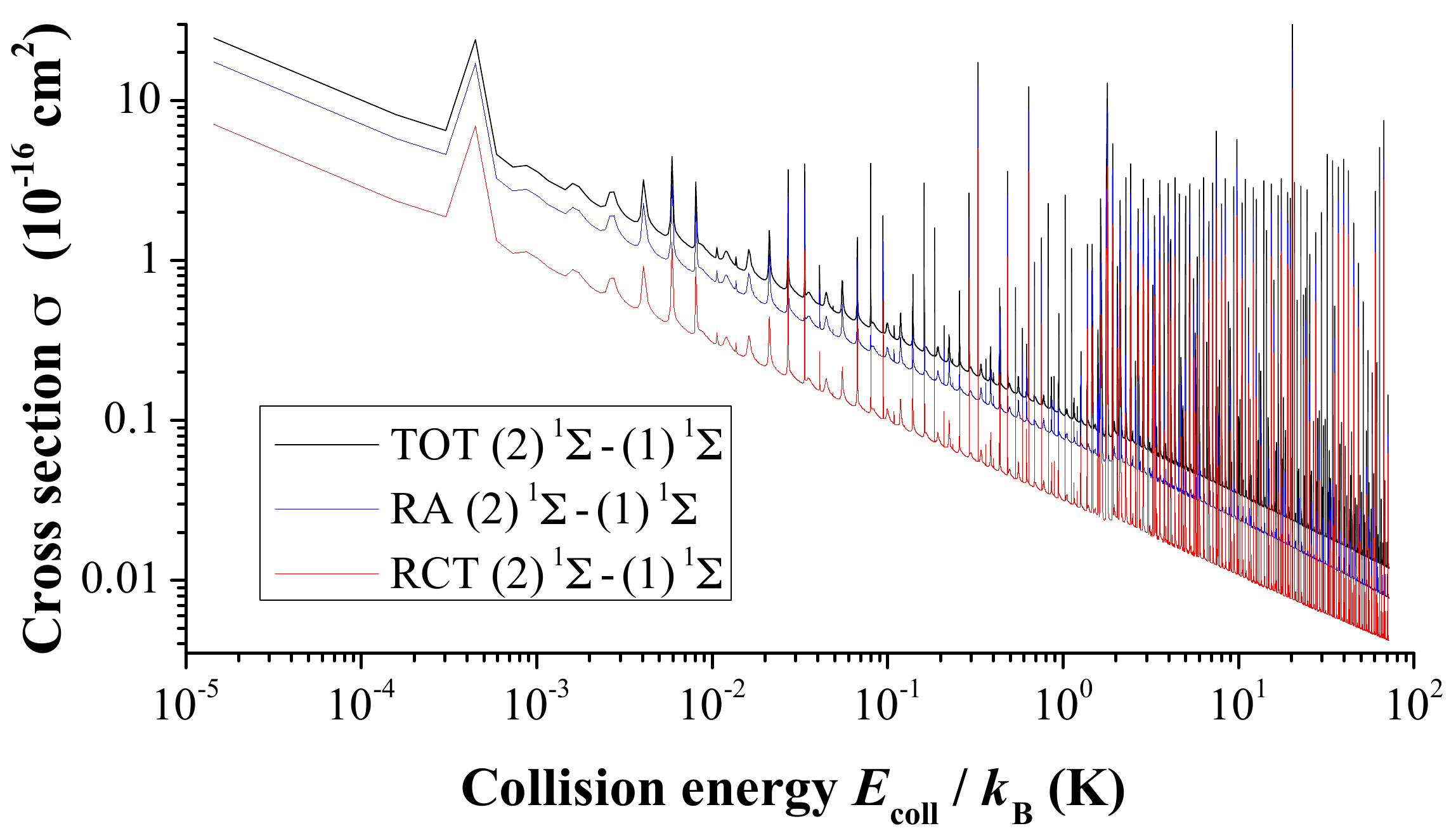,width=0.8\columnwidth}
\end{center}
\caption{\label{csfig} Theoretical cross sections for radiative charge transfer (RCT) and radiative association (RA) from the $(2)^1\Sigma^+$ molecular state associated with the Ca$^+(4s)$+Rb$(5s)$ entrance channel. TOT denotes the sum of both processes.}
\end{figure}

As already discussed in paper I, the Rb$^+$+Ca($4s4p$, $^3P$) asymptote is located 352~cm$^{-1}$ below the entrance channel. Note that the energy spread of the Ca($4s4p$, $^3P$) fine structure manifold (158~cm$^{-1}$) is smaller than this energy difference, and thus does not modify the order of the dissociation limits of the system. The two circles drawn in Fig. \ref{pec}(a) suggest that the $(2)^1\Sigma ^+$ state is coupled by spin-orbit (SO) interaction to the $(1)^3\Pi$ state leading to non-radiative charge transfer (NRCT). Gianturco and coworkers \cite{tacconi11a,belyaev12a} have shown that this process has a significant cross section at low energies. We may thus expect that NRCT competes with the radiative processes characterized in the present work (see Sec. \ref{pc} below). In principle, RA is possible from the $(1)^3\Pi$ state towards the $(1)^3\Sigma ^+$ state with a large TDM (dashed-dotted blue line in Fig. \ref{pec} (b)), but because of their small energy difference the relevant cross section is small in comparison to the singlet states.

The cross sections for the radiative processes between the lowest $^1\Sigma^+$ states are displayed in Fig. \ref{csfig}. The RA cross sections are predicted to be about 2.5 times larger than the RCT cross sections over the energy interval studied. Both, the RA and RCT cross sections show pronounced shape resonances reflecting the dynamic trapping of the wavefunction behind the centrifugal barrier at well-defined collision energies. The positions and intensities of the features in both the RA and RCT channels are identical, highlighting the fact that the resonance behaviour is primarily governed by the properties of the long-range and centrifugal potentials and not the details of the short-range dynamics. The pattern of shape resonances appears similar to the one found for the NRCT process in the same entrance channel in Ref. \cite{belyaev12a}. 

\subsection{Product branching ratios}
\label{pbr}

The population in the excited states of Ca$^+$ and Rb and therefore the contribution from excited reaction channels are critically dependent on the detunings of the Ca$^+$ and Rb cooling lasers. In paper I \cite{hall11a}, we characterized the reaction rates for the relevant excited reaction channels. Here, we study the product branching ratios using RE mass spectrometry. RE mass spectra were recorded under different Ca$^+$ laser-cooling conditions and resulting in varying contributions of the excited Ca$^+[(4p)~^2P_{1/2}]$/$[(3d)~^2D_{3/2}]$ + Rb$[(5s)~^2S_{1/2}]$ reaction channels. By contrast, adjusting the detuning of the Rb cooling laser did not have any measurable effect. This result can be rationalized first by the isolation of the molecular curves connecting to the Rb$(5p)$+Ca$^+(4s)$ asymptote which do not cross with any charge-transfer curves hence precluding NRCT, and second by their repulsive character at short range which lead to small Franck-Condon factors and hence a small radiative transition probability to the lower-lying molecular states (see Fig. 4 of paper I).

\begin{figure}[t]
\begin{center}
\epsfig{file=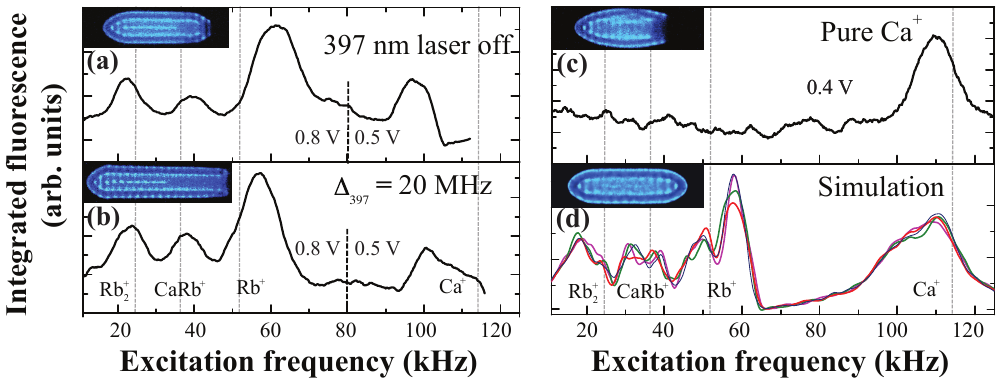,width=\textwidth}
\end{center}
\caption{\label{ms} Resonant-excitation mass spectra of Coulomb crystals: (a) after reaction without laser cooling of Ca$^+$, (b) after reaction with laser cooling both species, (c) pure Ca$^+$ crystal before reaction, (d) molecular-dynamics simulation of the spectrum in (b). The coloured traces represent different simulations performed with slightly varying initial conditions. See text for details.}
\end{figure}

Fig. \ref{ms} (a) shows a RE mass spectrum recorded after a reaction in which the 397~nm cooling laser was switched off after immersing the Coulomb crystal into the Rb cloud so that the Ca$^+$ population was entirely confined to the $(4s)~^2S_{1/2}$ ground state. As the ions were not laser cooled over the course of the reaction, no fluorescence images could be obtained. It is possible that the Coulomb crystal melted during that period and the reaction occurred with a diffuse ion cloud for which the overlap with the ensemble of ultracold atoms and the ion kinetic energies were not well defined. Under these conditions, the rate constant was determined to be $k_s=2\times10^{-12}$~cm$^3$s$^{-1}$. Given the uncertainty of the experimental conditions, we regard this result as an order-of-magnitude estimate for the rate constant in the lowest channel. A rigorous upper bound $k_s\leq 3\times10^{-12}$~cm$^3$s$^{-1}$ was determined in paper I. 

Four resonances can be distinguished in the spectrum in Fig. \ref{ms} (a). Three of them correspond to the product ions Rb$_2^+$, CaRb$^+$, Rb$^+$, one of them to the remaining Ca$^+$ ions in the crystal. Note that the peak maxima are slightly shifted from the predicted single-ion resonance frequencies (dashed lines) because of interspecies-coupling effects in the Coulomb crystals \cite{roth07a}. Moreover, the excitation amplitude has been reduced from 0.8~V to 0.5~V in the region of the strong Ca$^+$ resonance in order to prevent a melting of the crystal during the scan. CaRb$^+$ and Rb$^+$ are the products of the NRCT, RCT and RA processes discussed above. As already explained in paper I \cite{hall11a}, Rb$_2^+$ is likely the product of a consecutive reaction of CaRb$^+$ with Rb occurring in the hybrid trap. In principle, the generation of Rb$_2^+$ by photoionization or charge transfer with Rb$_2$ formed by photoassociation of Rb by the 780~nm cooling lasers is possible as well. However, Gabbanini et al. \cite{gabbanini00a} showed that the formation of $^{87}$Rb$_2$ molecules in a MOT is inefficient and governed by three-body recombination processes which can be neglected at the low Rb number densities used in the present experiment. Another possible route leading to Rb$_2^+$ is RA of Rb$^+$ product ions with excited Rb atoms in the MOT. Further studies to assess the possible role of this mechanism are currently in progress.

Fig. \ref{ms} (b) shows a spectrum obtained at the usual 397~nm-laser detuning  (20~MHz) used in the present experiments. These conditions correspond to Ca$^+$ excited-state populations $(p_p=12\%$ and $p_d=13\%$ where $p_p$ and $p_d$ denote the populations in the  $(4p)~^2P_{1/2}$ and $(3d)~^2D_{3/2}$ states, respectively. The similarity between the spectra in Figs. \ref{ms} (a) and (b) indicates that the product branching ratios are similar under both experimental conditions. For comparison, panel (c) shows the spectrum obtained for a pure Ca$^+$ crystal before reaction.

The relative contributions of the reaction channels to the product distribution can be estimated from the channel-specific rate constants and level populations. In paper I \cite{hall11a}, the rate coefficients for reactions in the Ca$^+~(3d)$ and $(4p)$ channels were determined to be $k_{d}\leq3\times10^{-12}$~cm$^3$s$^{-1}$ and $k_p=1.5(6)\times10^{-10}$~cm$^3$s$^{-1}$, respectively. As the product of the rate constant with the population for the Ca$^+(3d)$ channel is at least an order of magnitude smaller than for the other two channels, we neglect its contribution. The effective rate constants for the Ca$^+(4s)$ and $(4p)$ channels are calculated to be $k_s^\prime=(1/2)(1+p_s)k_s$ and $k_p^\prime=(1/2)p_pk_p$, respectively (see the kinetic model discussed in paper I \cite{hall11a}). For the conditions corresponding to Fig. \ref{ms} (b), one obtains $k_s^\prime\approx 1.7\times10^{-12}$~cm$^3$s$^{-1}$ and $k_p^\prime=9.0\times10^{-12}$~cm$^3$s$^{-1}$, implying that the Ca$^+(4s)$ and $(4p)$ channels contribute with a ratio of $\approx$1:5 to the product yields observed in the RE spectra. However, considering that all of the three processes NRCT, RCT and RA are also expected to occur in the excited Ca$^+(4p)$ channel with significant efficiencies as argued in paper I, the similarities between the spectra in Figs. \ref{ms} (a) and (b) are not necessarily surprising.

The RE spectra were further analyzed with the help of MD simulations shown in panel (d). A crystal composition of 170 Ca$^+$, 120 Rb$^+$, 39 CaRb$^+$ and 52 Rb$_2^+$ ions was found to best reproduce the experimental peak intensities. The different simulations shown in panel (d) were performed using slightly different initial conditions to check the robustness of the final result. However, because of the approximations inherent in the simulation procedure (see section \ref{massspec}) and the sensitivity of the experimental spectra to the specific experimental conditions, the comparison between experiment and simulation can only be expected to provide a rough estimate of the relative numbers of the ion species in the crystal. 

\subsection{Collision-energy dependence of reaction rates}
\label{pc}

\begin{figure*}[t]
\begin{center}
\epsfig{file=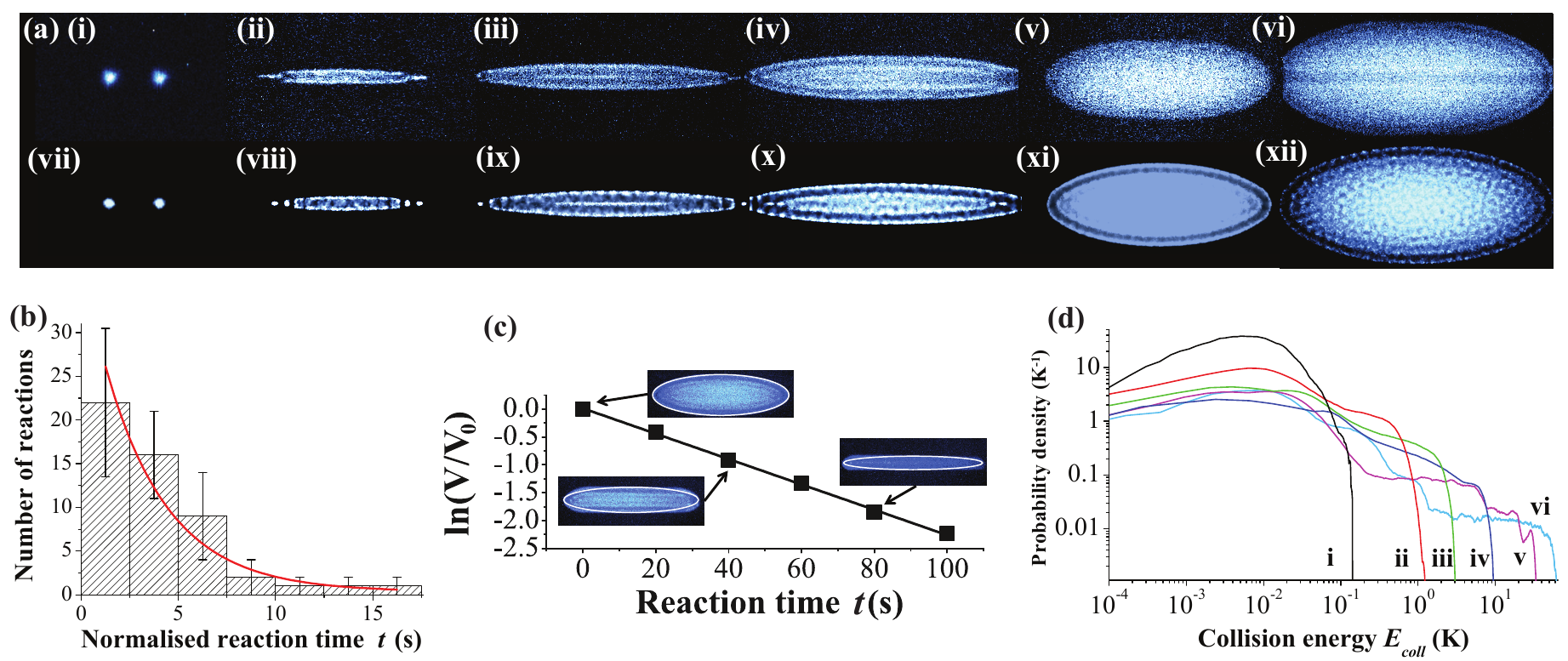,width=\textwidth}
\end{center}
\caption{\label{ecol} Variation of collision energies by reacting Ca$^+$ Coulomb crystals of different sizes and shapes containing ions with different degrees of micromotion excitation. (a) (i)-(vi) experimental images, (vii)-(xii) MD simulations. (b) Determination of the rate constant from a series of single-ion reaction experiments enabling the lowest collision energies achieved in the present study. (c) Illustration of a fit of the time-dependent Coulomb-crystal volume $V$ to a pseudo-first order rate law to determine reaction rate constants for large crystals. (d) Collision-energy distributions for the reaction experiments with the Coulomb crystals (a) (i)-(vi) derived from the ion trajectories obtained in the simulations (a) (vii)-(xii). See text for details.}
\end{figure*}

The dependence of the rate constant on the collision energy represents an important dynamical characteristic of collision systems. In the present experiments, the energies of collisions between Ca$^+$ and Rb were varied using two different approaches. Rate constants at medium to high average collision energies in the range $\langle E_{\text{coll}}\rangle /k_{\text{B}}\approx 200$~mK-20~K were obtained by reacting Coulomb crystals of varying size and shape as shown in Figs. \ref{ecol} (a) (ii)-(vi). Because the kinetic energy of the ions depends on their position in the RF field of the trap, crystals of different size and shape exhibit different ion-kinetic-energy distributions (Fig. \ref{ecol} (d)) and therefore different {\em average} ion kinetic energies \cite{bell09a}. Reaction rates were determined by a fit of the time-dependent volume $V$ of the Ca$^+$ Coulomb crystals to a pseudo-first-order rate law $\ln(V/V_0)=-k^\prime t$ \cite{willitsch08a}, where $V_0$ is the initial volume and $t$ is the reaction time (Fig. \ref{ecol} (c)). Second-order rate constants $k$ were obtained by dividing the pseudo-first order constants $k^\prime$ by the average Rb atom density in the volume of the crystal \cite{eberle12a}.

The lowest collision energies $\langle E_{\text{coll}}\rangle /k_{\text{B}}\approx20$~mK were achieved in a series of experiments in which single Ca$^+$ ions were reacted from a string of two Ca$^+$ ions localized on the central trap axis. The rate constant at this collision energy was obtained from a pseudo-first order fit of the time to reaction observed in a total of 49 single-ion reaction experiments as shown in Fig. \ref{ecol} (b) (see also Ref. \cite{staanum08a}).

Fig. \ref{ecol} (d) shows a double-logarithmic representation of the collision-energy distributions of Rb atoms with the experimental Coulomb crystals (i)-(vi) derived from the MD simulations Fig. \ref{ecol} (a) (vii)-(xii). The collision energy distributions were calculated from the kinetic energies obtained from the ion trajectories. The kinetic energy of the Rb atoms is at least two orders of magnitude lower and was neglected. As can be seen from Fig. \ref{ecol} (d), the collision-energy distributions of the crystals are highly non-Maxwellian reflecting the non-thermal character of the micromotion of the ions (see also Ref.  \cite{bell09a}). The narrowest collision-energy distribution is obtained for the 2-ion string (i), whereas for increasingly larger crystals distributions with a broader width are calculated. For the crystal (vi), the average collision energy amounts to 16~K with peak energies as high as 60~K.

\begin{figure}[t]
\begin{center}
\epsfig{file=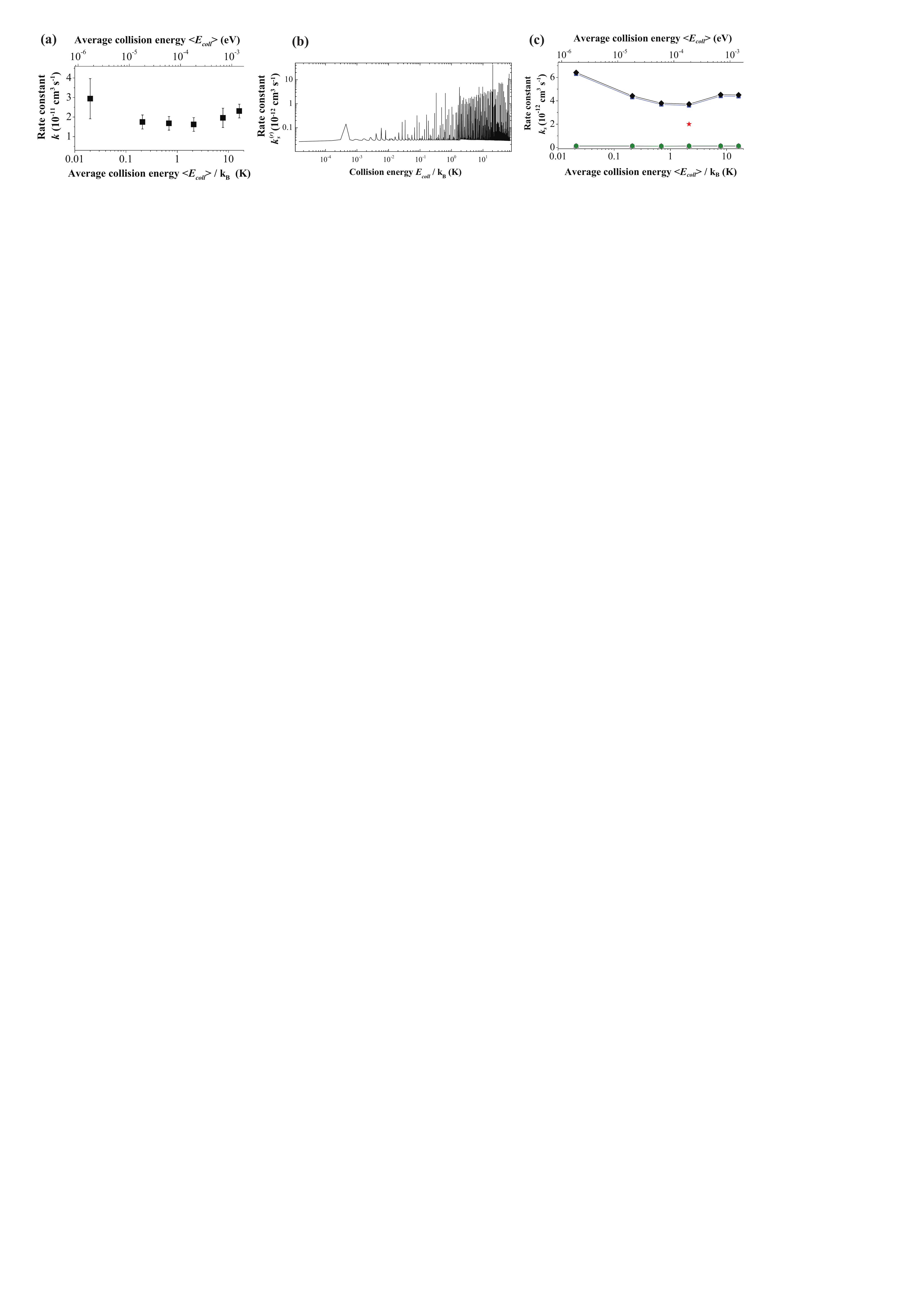,width=\textwidth}
\end{center}
\caption{\label{etcomp} (a) Rate constant $k$ as a function of the average collision energy for Coulomb crystals exhibiting the shapes displayed in Fig. \ref{ecol} (a). (b) Theoretical total radiative rate constant $k_s^{(r)}$ from the ground state entrance channel. (c) Theoretical rate constant $k_s$ for the Ca$^+(4s)$+Rb$(5s)$ channel as a function of the collision energy. Green circles: total radiative rate constants, violet triangles: total radiative rate constants without resonances, blue squares: non-radiative rate constants calculated from the cross sections reported in Ref. \cite{tacconi11a}, black diamonds: summed radiative and non-radiative rate constants. Red star: experimental result at $\langle E_{\text{coll}}\rangle/k_{\text{B}}=$2~K.}
\end{figure}

The experimental rate constants determined for crystals with shapes as shown in Figs. \ref{ecol} (a) (i)-(vi) are displayed in Fig. \ref{etcomp} (a). Note that the rate constants have been normalized with respect to the overlap of the Coulomb crystals with the Rb cloud and therefore differ slightly, but not significantly, from the unnormalized values reported in paper I. The rate constant was found to be essentially constant within the error boundaries in the interval $\langle E_{\text{coll}}\rangle/k_{\text{B}}=$20~mK-20~K.

Fig. \ref{etcomp} (b) shows a double-logarithmic representation of the total theoretical radiative rate constant $k_s^{(r)}$ in the lowest reaction channel in the energy interval 10$~\mu$K$-$70~K obtained by multiplying the total radiative cross sections (Fig. \ref{csfig}) by the velocity of the collision partners. Disregarding the narrow shape resonances, the rate constant is essentially constant with collision energy. 

This energy dependence can be understood in terms of a classical model for the reaction dynamics. Note that the base line of the cross-section curves in Fig. \ref{csfig} varies with energy as $\epsilon_i^{-1/2}$. This behaviour is predicted by the classical Langevin model of ion-neutral capture \cite{gioumousis58a, gao11a}. Indeed, the computed cross sections can be recovered in good approximation if the Langevin cross section $\sigma^{\text{(L)}}$ is multiplied by a constant opacity factor $P$ to account for a non-unit reaction probability at short range:
\begin{equation}
\sigma=P\sigma^{\text{(L)}}=P\pi q \sqrt{\frac{2\alpha}{\epsilon_i}}. \label{langevin}
\end{equation}
In Eq. (\ref{langevin}), $\alpha$ stands for the isotropic polarizability of the neutral reaction partner and $q$ is the charge of the ion. The associated rate constant $k=\sigma v\propto \sigma \epsilon_i^{1/2}$ is constant with energy. The fact that the energy dependence of the cross sections can be recovered with a classical model is not surprising considering the still large number of partial waves which contribute to the collisions even at the low energies reached in the present study (from $J\leq22$ at $E_{\text{coll}}\approx10$~mK to $J\leq 200$ at $E_{\text{coll}}\approx50$~K). The quantum character of the collisions primarily manifests itself in the occurrence of the shape resonances. Refs. \cite{turulski94a, dashevskaya03a, nikitin05a, roudjane07a, hall12a} include more detailed discussions on the validity of ion-neutral capture-models at low energies.

The classical behaviour for the energy dependence can also be recovered from the quantum-mechanical theory described in section \ref{scattcalc}. The transition-dipole moment function $D(R)$ in Eqs. (\ref{csrct}) and (\ref{csra}) is non-vanishing only for short internuclear distances $R$ (see Fig. \ref{pec} (b)). In this region, the amplitude of the continuum wave function is nearly independent of the collision energy $\epsilon_i$ which is small compared to the well depth in the entrance channel. The matrix element in Eq. (\ref{dipcc}) is thus nearly constant as a function of $\epsilon_i$. Ignoring the shape resonances, we find that the TDMs in Eq. (\ref{csrct}) and (\ref{csra}) are also nearly independent of the total angular momentum $J$ as long as $J<J_{\text{max}}$, where $J_{\text{max}}$ is the maximum value of $J$ for which the collision energy exceeds the height of the centrifugal barrier. This finding reflects classical behaviour, i.e., the collision partners can approach to short internuclear distances to react only if the incident energy is large enough to surmount the centrifugal barrier. 

Consequently, the TDM can be factored out of the sums in Eqs. (\ref{csrct}) and (\ref{csra}). The energy dependence of the cross section then results from the summation over $J$ up to the maximum allowed value $J_{\text{max}}$ yielding $\sigma \propto J_{\text{max}}^2/k_i^2 \equiv b^2_{\text{max}}$, where $b_{\text{max}}$ is the maximum classical impact parameter of the collision at the energy $\epsilon_i$. $b_{\text{max}}\propto\epsilon_i^{-1/4}$ for the present long-range interaction potential scaling as $V(R)\propto R^{-4}$ \cite{levine05a}. Thus, $\sigma\propto \epsilon_i^{-1/2}$ as in the classical Langevin model and the rate constant is independent of the collision energy. By contrast, for entrance channels exhibiting long-range potentials with a different $R$ dependence, the rate constant is expected to be dependent on the energy, see, e.g., the recent results on N$_2^+(^2\Sigma^+_g)$+Rb$(5p~^2P_{3/2})$ collisions reported in Ref. \cite{hall12a}.

We note that the same arguments are expected to hold for the NRCT processes, and indeed the NRCT cross sections reported in Ref. \cite{tacconi11a} approximately scale as $\sigma\propto \epsilon^{-1/2}$. Therefore, the dynamics of the reactive collisions can be separated into two contributions: long-range classical capture which governs the relative energy dependence of the cross sections, and the short-range dynamics of the different channels which determines their magnitudes.

Fig. \ref{etcomp} (c) shows the rate constants $k_s$ for the lowest collision channel predicted for the present experimental conditions. The rate constants were calculated from the theoretical cross sections Fig. \ref{csfig} using the relationship \cite{gerlich92a}
\begin{equation}
k_s=\int_0^\infty \sigma(v)v\rho(v)dv, \label{conv}
\end{equation}
where $v$ is the relative velocity of the collision partners and $\rho(v)$ is the normalized distribution of relative velocities calculated from the energy distributions displayed in Fig. \ref{ecol} (d). The green circles and violet triangles represent the total radiative rate constants including and excluding the shape resonances, respectively. The differences are small (the symbols are almost perfectly superimposed on the scale of the figure) indicating that the overall contribution of the resonances to the averaged rate is only slight. The blue squares indicate the rate constants calculated from the NRCT cross sections for the lowest reaction channel taken from Ref. \cite{tacconi11a}. The black diamonds represent the summed radiative and non-radiative rate constants and the red star indicates the experimental estimate for the lowest reaction channel at $\langle E_{\text{coll}}\rangle/k_{\text{B}}$=2~K.

At $\langle E_{\text{coll}}\rangle/k_{\text{B}}$=2~K, the relative rate constants for the radiative and non-radiative processes differ by a factor of $\approx40$. Whether this difference is significant cannot be ascertained at the moment, considering that different potential curves and numerical approximations were used for the calculation of the radiative processes in the present study and NRCT in Ref. \cite{tacconi11a}. Indeed, the analysis of the product branching ratios in Sec. \ref{pbr} suggests that the ratio of the molecule-formation vs. charge-transfer rates is $\approx 1:3$, implying that RA processes are more prominent relative to NRCT than indicated in Fig. \ref{etcomp} (c). A future unified theoretical treatment of the radiative and non-radiative processes is required to address this question conclusively.  

The data in Fig. \ref{etcomp} (a) was taken under similar conditions as the RE mass spectrum in Fig. \ref{ms} (b), i.e., the Ca$^+(4s)$+Rb$(5s)$ and Ca$^+(4p)$+Rb$(5s)$ channels contribute with a ratio of $\approx 1:5$. The lowest channel thus only represents a minor contribution to the observed rate constant $k$ so that the magnitudes of the rate constants in Figs. \ref{etcomp} (a) and (c) are not directly comparable. However, the predicted relative energy dependence of the rate constant $k_s$ for the lowest channel (Fig. \ref{etcomp} (c)) closely resembles the energy profile of the observed rate constant $k$ (Fig. \ref{etcomp} (a)), suggesting that the energy dependence of the rates is similar for the Ca$^+(4s)$+Rb$(5s)$ and Ca$^+(4p)$+Rb$(5s)$ channels. Indeed, both channels exhibit long-range potentials scaling as $R^{-4}$ and are thus expected to show the same scaling behaviour of the reaction rate with energy (see above).

\section{Summary and conclusions}
\label{conclusions}

In the present study we have characterized the dynamics of chemical reactions between laser-cooled Ca$^+$ ions and Rb atoms in an ion-atom hybrid trap. For the cold scattering experiments performed here, we show that the observed reaction rates can be explained by two main features. First, the energy variation of the reaction rate constants is ruled by universal behaviour governed by the the long-range charge-induced dipole interaction between the reaction partners in line with the classical Langevin model. Second, their magnitude is determined by short-range non-adiabatic and radiative couplings which induce chemical changes specific to the relevant reaction channel. The chemistry in the Ca$^+$+Rb system is dominated by several competing processes, i.e., non-radiative as well as radiative charge transfer and radiative association leading to the formation of CaRb+ molecular ions. The molecular ions are sympathetically cooled by the atomic ions in the Coulomb crystal and may be used for further chemical studies in the cold domain. 

We expect this picture to be widely applicable to cold reactive collisions between ions and atoms in a regime in which the collision can still be treated classically (i.e., many partial waves contribute), but the collision energy is low enough so that the short-range coupling matrix elements responsible for the reactive processes are approximately constant with energy. The latter is usually the case when the coupling is localized at short internuclear distances in a sufficiently deep potential well. We note that similar energy dependences of the rate constants have also been observed in Ba$^+$+Rb \cite{hall13b} and Yb$^+$+Ca \cite{rellergert11a}. These observations can be rationalized using the present model.

The quantum character of the collisions manifests itself in the occurrence of narrow shape resonances at well-defined collision energies, which, however, have little impact on the ensemble-averaged rates for the present system and conditions. However, for collision systems such as B$^{5+}$+H  and N$^{3+}$+H which are considerably lighter than the one studied here, the width of the resonances is considerably increased so that a classical treatment of the collision process is inappropriate \cite{barragan10a, roudjane07a}. 

The present study has established an order-of-magnitude agreement between experiment and theory. To enable more stringent tests of the  models for cold reactive scattering, both a unified theoretical treatment of all reactive processes and more accurate experiments for the lowest reaction channel have to be established. Work in this direction is currently in progress.

\section*{Acknowledgements}

We acknowledge support from the Swiss National Science Foundation (grants nr. PP0022\_118921 and PP00P2\_140834) and the COST Action MP1001 ''Ion Traps for Tomorrow's Applications''. We thank Dr Alexander Gingell for making available his program for the 8-level OBE treatment of Ca$^+$. 

\appendix

\section{Details of experimental procedures}
\subsection{Axialization of the Coulomb crystals}
\label{ax}

\begin{figure}[t]
\begin{center}
\epsfig{file=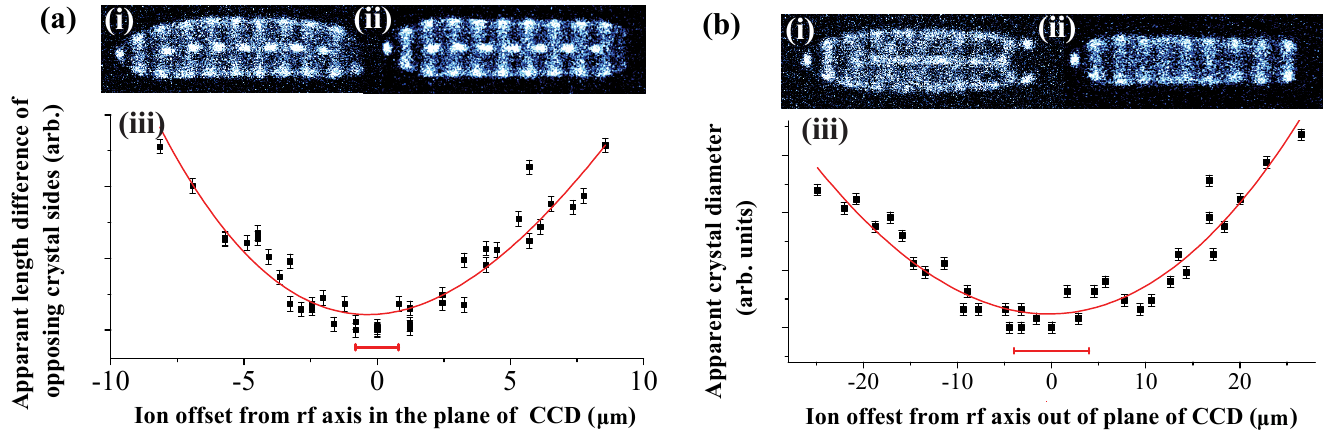,width=\textwidth}
\end{center}
\caption{\label{cc} (a) Alignment of Coulomb crystals onto the central trap axis (``axialization'') in the image plane of the CCD camera: (i) crystal shifted off axis, (ii) crystal optimally positioned on the axis, (iii) optimal axialization if the difference of the length of the opposing straight crystal edges is minimized. (b) Axialization along the line of sight of the CCD camera: (i) crystal shifted off axis, (ii) crystal optimally positioned on the axis, (iii) optimal axialization is achieved if the apparent vertical crystal diameter is minimized. See text for details.}
\end{figure}

Axialisation of the Coulomb crystals was achieved by taking advantage of the mass-dependence of the effective trapping potential for different ion species in a bi-component Coulomb crystal \cite{willitsch08b, willitsch12a}. If the crystal is perfectly centred on the trap axis, heavy sympathetically-cooled ions arrange in concentric shells around a core of lighter laser-cooled ions \cite{bell09a, willitsch08b, willitsch12a}. In the fluorescence images, this effect manifests itself as a symmetric distribution of dark sympathetically-cooled ions at the edges of the crystal leading to a characteristic flattening of the central Ca$^+$ core, see Fig. \ref{cc} (a) (ii). By contrast, if the crystals are shifted off-axis by stray electric fields, the radial symmetry of the trapping potential is broken resulting in an asymmetric segregation of the ion species and deformed Ca$^+$ crystal shapes, see Fig. \ref{cc} (a) (i). In this case, axialisation of the crystals can be achieved by the application of appropriate static compensation fields to different segments of the trap electrodes. Axialization in the plane of the images is achieved if an equal distribution of sympathetically-cooled ions is obtained on both sides of the crystal image, see Fig. \ref{cc} (a) (ii). A practical figure of merit for axialization in the plane of the image is the minimization of the difference of the length of the opposing straight crystal edges, see Fig. \ref{cc} (a) (iii).

If the crystal is shifted  off axis perpendicular to the plane of the images (along the line of sight of the camera), the sympathetically-cooled ions localize either in front of or behind the image plane, manifesting itself as an increase of the apparent diameter of the core of laser-cooled ions in the images in comparison to a crystal perfectly axialized along the line of sight, see Fig. \ref{cc} (b) (i) vs. (ii). In this case, axialization is achieved by moving the crystal using static fields until the apparent diameter of the Ca$^+$ ion core is minimized, see Fig. \ref{cc} (b) (ii).

With these simple methods, the ion crystals could be axialized with an accuracy of $\approx 1~\mu$m and $\approx 4~\mu$m in the directions parallel and perpendicular to the image plane, respectively. For a string of ions, the position uncertainties translate into a micromotion-related uncertainty in the ion kinetic energies of $E_{\text{kin}}/k_{\text{B}}\approx1$~mK and  $\approx20$~mK in and out of plane of the images, respectively. During the experiments, the axialization of the Coulomb crystals was checked periodically to account for time-varying stray electric fields in the apparatus.

\subsection{Spatial alignment of the cloud of ultracold atoms with the Coulomb crystals}
\label{apmot}

\begin{figure}[t]
\begin{center}
\epsfig{file=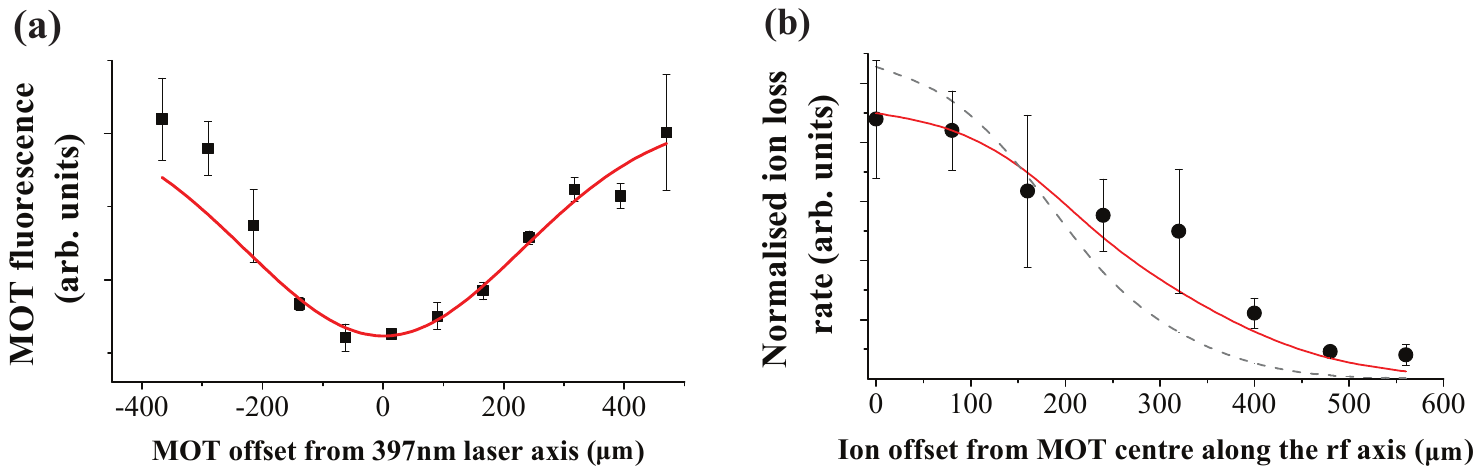,width=\textwidth}
\end{center}
\caption{\label{mot} Alignment of the centres of the MOT and the ion trap along the line of sight of the CCD camera: (a) MOT fluorescence as a function of displacement from the 397 nm Ca$^+$ cooling laser beam, (b) Ca$^+$ loss rate as a function of the offset of the Coulomb crystal from the centre of the MOT. Dashed line: prediction using a static overlap model, red solid line: MOT radius scaled by a factor 1.4. See text for details. }
\end{figure}

For the correct determination of reaction rates, an accurate alignment of the position of the cloud of ultracold atoms with respect to the ion Coulomb crystal is indispensable. While the alignment of both species in the image plane is straight forward (see Fig. \ref{setup}), ensuring a perfect overlap along the line of sight of the camera is more challenging. In the present experiments, the 397~nm laser beam was aligned for maximum overlap with the ion crystal to maximize the Ca$^+$ laser-cooling efficiency and fluorescence intensity. When all laser beams were simultaneously admitted to the trap, the 397~nm laser beam served to ionize Rb atoms from the MOT, leading to a minimum in the Rb atom fluorescence when the MOT was optimally overlapped with the ion crystal as shown in Fig. \ref{mot} (a).

Fig. \ref{mot} (b) shows the normalized reaction rate of Ca$^+$ ions from the Coulomb crystal as a function of the displacement of the MOT from the center of the ion trap. The reaction rate decreases with the offset of the MOT from the Coulomb crystal, reflecting the decreasing degree of overlap with the cloud of ultracold Rb atoms. The dashed line shows a theoretical prediction of the data taking into account a static geometric overlap of the two species. The theoretically predicted rates fall off faster with increasing MOT displacement than the measured ones. This result suggests that the diameter of the cloud of ultracold atoms is effectively larger than the one assumed in the static overlap model which was determined from atom fluorescence images taken at the beginning of each measurement. Indeed, over the duration of the experiments (several minutes), the position and shape of the atom cloud fluctuates slightly because of instabilities in the 780~nm-laser intensity and wavelength, so that the atoms effectively sample a larger volume over that period than can be inferred from a static image only. Scaling the MOT radius by a factor 1.4 to account for the increase in effective Rb cloud volume yielded a good representation of the experimental results (red line in Fig. \ref{mot} (b)).

\subsection{Determination of Ca$^+$ level populations}
\label{appop}

\begin{figure}[t]
\begin{center}
\epsfig{file=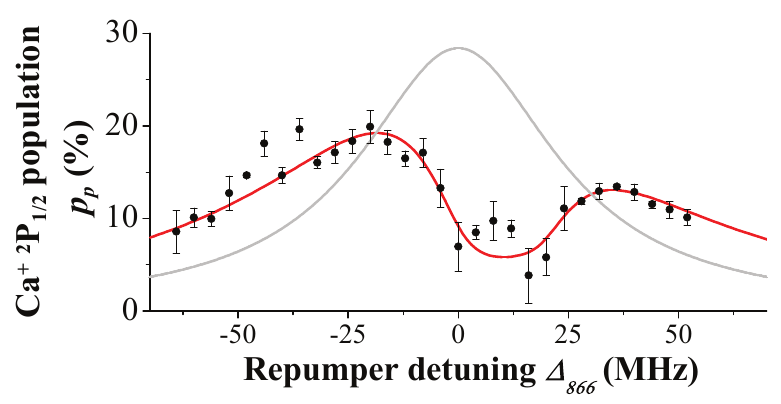,width=0.6\textwidth}
\end{center}
\caption{\label{pop} Ca$^+~(4p)~^2P_{1/2}$ level population as a function of the repumper-laser detuning $\Delta_{866}$. Red line: prediction from 8-level optical-Bloch-equation model, grey line: prediction from rate equation model. See text for details.}
\end{figure}

Fig. \ref{pop} shows the Ca$^+~^2P_{1/2}$ level populations as a function of the 866~nm repumper-laser detuning. The data points represent populations derived from photon-counting measurements of single spatially resolved Ca$^+$ ions using the CCD camera. The populations have been calculated from the number of 397~nm photons emitted on the $(4p)~^2P_{1/2} \rightarrow (4s)~^2S_{1/2}$ transition taking into account the natural lifetime of the $^2P_{1/2}$ state, the solid angle of the fluorescence captured by the camera, the quantum yield of the CCD chip, losses at the collection optics, the gain factor of the camera's image intensifier and the branching ratio of fluorescent decay to the $^2D_{3/2}$ state. The populations show a marked minimum at equal detunings of the 397~nm and 866~nm lasers ($\approx15$~MHz) resulting from coherent population trapping (CPT) \cite{arimondo96a}. The red solid curve shows the populations calculated using an 8-level Bloch-equation model including the effects of applied magnetic fields \cite{gingell10a}. The magnetic field was approximated to be homogeneous across the position of the ions and the only free parameters in the model were its strength and angle with respect to the laser polarization axis. These parameters were adjusted to reproduce the width and depth of the CPT resonance. The solid grey curve represents the populations calculated with an Einstein-type rate equation model \cite{foot05a} which does not account for coherent effects and is therefore unable to reproduce the dark resonance. Note that while the populations calculated with the rate-equation model deviate substantially from the measured ones in the vicinity of the dark resonance, they agree within better than a factor of 2 at large detunings $|\Delta_{866}|\geq 25$~MHz typically used in the present reaction experiments.

\bibliographystyle{tMPH}

\end{document}